\documentclass[pdflatex,sn-mathphys-ay]{sn-jnl}

\usepackage{graphicx}%
\usepackage{multirow}%
\usepackage{amsmath,amssymb,amsfonts}%
\usepackage{amsthm}%
\usepackage{mathrsfs}%
\usepackage[title]{appendix}%
\usepackage{xcolor}%
\usepackage{textcomp}%
\usepackage{manyfoot}%
\usepackage{booktabs}%
\usepackage{listings}%
\usepackage{float}
\usepackage{cleveref}
\usepackage{graphicx}
\usepackage{hyperref}
\usepackage{adjustbox}
\usepackage{subcaption}
\usepackage{caption}
\usepackage{geometry}
\usepackage{multirow}
\usepackage{enumerate}
\usepackage{array}
\usepackage{xurl}
\usepackage{algorithm}
\usepackage{algpseudocode}

\theoremstyle{thmstyleone}%
\theoremstyle{thmstyletwo}%

\theoremstyle{thmstylethree}%

\begin{document}

\title[Analyzing the Temporal Factors for Anxiety and Depression]{Analyzing the Temporal Factors for Anxiety and Depression Symptoms with the Rashomon Perspective}


\author[1]{\fnm{Mustafa} \sur{Cavus}}

\author[2]{\fnm{Przemys\l{}aw} \sur{Biecek}}
 
\author[3]{\fnm{Julian} \sur{Tejada}}

\author*[4]{\fnm{Fernando} \sur{Marmolejo-Ramos}}\email{fernando.marmolejoramos@flinders.edu.au}

\author[3]{\fnm{Andre} \sur{Faro}}

\affil[1]{\orgdiv{Department of Statistics}, \orgname{Eskisehir Technical University}, \orgaddress{\country{Türkiye}}}

\affil[2]{\orgdiv{Department of CAD/CAM Systems Design and Computer-Aided Medicine. Faculty of Mathematics and Information Sciences}, \orgname{Warsaw University of Technology}, \orgaddress{\state{Mazowieckie}, \country{Poland}}}

\affil[3]{\orgdiv{Department of Psychology}, \orgname{Federal University of Sergipe}, \orgaddress{\state{Sergipe}, \country{Brazil}}}

\affil[4]{\orgdiv{Collegue of Education, Psychology, and Social Work}, \orgname{Flinders University}, \orgaddress{\state{S.A.}, \country{Australia}}}


\abstract{This paper introduces a new modeling perspective in the public mental health domain to provide a robust interpretation of the relations between anxiety and depression, and the demographic and temporal factors. This perspective particularly leverages the Rashomon Effect, where multiple models exhibit similar predictive performance but rely on diverse internal structures. Instead of considering these multiple models, choosing a single best model risks masking alternative narratives embedded in the data. To address this, we employed this perspective in the interpretation of a large-scale psychological dataset, specifically focusing on the Patient Health Questionnaire-4. We use a random forest model combined with partial dependence profiles to rigorously assess the robustness and stability of predictive relationships across the resulting Rashomon set, which consists of multiple models that exhibit similar predictive performance. Our findings confirm that demographic variables \texttt{age}, \texttt{sex}, and \texttt{education} lead to consistent structural shifts in anxiety and depression risk. Crucially, we identify significant temporal effects: risk probability demonstrates clear diurnal and circaseptan fluctuations, peaking during early morning hours. This work demonstrates the necessity of moving beyond the best model to analyze the entire Rashomon set. Our results highlight that the observed variability, particularly due to circadian and circaseptan rhythms, must be meticulously considered for robust interpretation in psychological screening. We advocate for a multiplicity-aware approach to enhance the stability and generalizability of ML-based conclusions in mental health research.}

\keywords{explainable artificial intelligence, partial dependence profile, Rashomon effect, public health, PHQ4, depression, anxiety}

\maketitle

\section{Introduction}\label{sec1}

Machine learning (ML) models are increasingly applied in public mental health to predict individual risk, monitor population-level trends, prioritize scalable digital interventions, and model complex, multidimensional relationships \citep{CruzGonzalez2025,DelPozoBanos2024,Islam2024,Iyortsuun2023,Sharma2025}. In such settings, multiple ML models—often trained on the same dataset—can achieve comparable predictive performance while relying on fundamentally different internal structures, feature dependencies, and decision rules. This phenomenon, known as the Rashomon effect, describes the existence of many distinct models that fit the same data equally well while offering different explanatory narratives \citep{breiman2001statistical}. This situation is particularly consequential in psychological screening and public mental health research, when equally accurate models may yield divergent clinical interpretations or policy recommendations by emphasizing different determinants. For instance, two models examining the determinants of depression prevalence in urban communities may attribute the disparities to different sources—one to socioeconomic inequality and access to care, another to social isolation or neighborhood-level stressors. The tendency to privilege a single model can therefore obscure alternative, yet equally plausible, explanations inherent in the data \citep{watson2024predictive}. This multiplicity is especially likely in health and psychological sciences, where data are frequently noisy, incomplete, or underspecified, and where measurement error or insufficiently expressive features limit the faithful representation of underlying phenomena \citep{rudin2024position}.

Therefore, model selection should go beyond accuracy and embrace the Rashomon set---the collection of the near-optimal models that yield a prediction error within a predefined small margin of the best performing model---as a space of epistemological exploration. Analyzing the similarities and divergences across these models allows researchers to assess the robustness, fairness, and interpretability of their conclusions \citep{rudin2025leo}. In this way, the Rashomon effect reveals not only the instability of ``the best model'' but also offers a lens through which multiple, equally plausible explanations can be surfaced \citep{cavus2025beyond}. To operationalize this epistemological exploration, modern techniques are required to systematically analyze the functional relationships captured by the Rashomon set. Framing this shift from single-model interpretation to multiplicity-aware analysis, we refer to this interpretation strategy as \textit{rashomonizing} the analysis. By this term, we denote the explicit incorporation of the Rashomon perspective into the interpretation of predictive models. While the Rashomon effect has been widely discussed as a descriptive phenomenon, the term \emph{rashomonize} is introduced here to emphasize its deliberate and systematic use as an interpretative procedure. In this paper, Partial Dependence Profiles (PDPs) \citep{friedman2001greedy} coupled with bootstrap confidence intervals allow us to visualize how the average prediction of the model changes across subgroups and variable ranges \citep{cavus2025beyond}. This approach directly confronts interpretation ambiguity by quantifying the stability of predictor effects—a necessity when dealing with complex, potentially high-stakes predictions like those in psychological screening.

Building on this epistemological foundation, the present study addresses the challenge of model multiplicity by applying a Rashomon approach to a large-scale mental health dataset. We specifically analyze scores from the Patient Health Questionnaire-4 (PHQ-4), an ultra-brief screening instrument for anxiety (GAD-2) and depression (PHQ-2) symptoms, which has been used for large-scale population monitoring in the United States \citep{kalu_role_2024} and Germany \citep{riepenhausen_coping_2023}, as well as in national and large online samples in Chile \citep{moreno_factor_2024} and Colombia \citep{sanabria-mazo_dimensionality_2024}, supporting its applicability for brief surveillance of combined depressive and anxiety symptom burden in diverse public health contexts. The dataset under consideration comprises over 34,000 records derived from adults who completed the PHQ-4. Employing a Random Forest model combined with bootstrap sampling, we generate PDPs for a set of demographic and temporal variables such as age, sex, education, and hour of day. Our methodology aims to not only identify key predictors of distress but also to rigorously assess the robustness and stability of these predictive relationships across the resulting Rashomon set. Finally, we discuss how the observed variabilities, particularly those related to circadian and circaseptan rhythms, highlight the necessity for a multiplicity-aware approach in interpreting psychological screening data. In summary, this paper makes three main contributions: 

\begin{enumerate}
    \item We operationalize the Rashomon effect in a public mental health context by explicitly analyzing a Rashomon set of Random Forest models rather than relying on a single best-performing model.
    \item We introduce a systematic framework that combines Partial Dependence Profiles with bootstrap confidence intervals, based on the Random Forest model under iterations, to quantify the stability and variability of predictor effects across near-optimal models, thereby addressing interpretation uncertainty induced by model multiplicity.
    \item We provide large-scale empirical evidence from PHQ-4 screening data demonstrating how demographic and temporal factors—including circadian and circaseptan patterns—exhibit heterogeneous yet robust effects within the Rashomon set, highlighting the practical implications of multiplicity-aware interpretation for psychological screening and population-level mental health monitoring.
\end{enumerate}

\noindent The remainder of this paper is structured as follows: Section~\ref{sec2} provides the necessary preliminaries on the Rashomon effect and the formal definition of the $\textit{Rashomon set}$. In Section~\ref{sec3}, we detail the dataset used, which consists of over 34,000 PHQ-4 scores, and outline the experimental design involving Random Forest modeling, bootstrap confidence intervals, and Partial Dependence Profiles. Section~\ref{sec4} presents the results, illustrating how demographic and temporal factors influence predicted risk for anxiety and depression symptoms. Finally, Section~\ref{sec5} offers a discussion of our findings, focusing on the implications of the Rashomon effect for robust interpretation in public mental health, and concludes the paper.

\section{Rashomon Perspective}\label{sec2}

The term \textit{Rashomon effect} originates from Akira Kurosawa’s 1950 film called Rashomon\footnote{https://en.wikipedia.org/wiki/Rashomon}, where different witnesses provide equally plausible but conflicting accounts of the same event. In machine learning, this concept was used to describe the inherent tendency of rich, complex, or high-dimensional datasets to support multiple models with similar predictive power \citep{breiman2001statistical}.

Traditionally, the Rashomon set is defined as the collection of models whose empirical loss is within a small margin $\varepsilon$ of the best-performing model. However, in this study, we operationalize the Rashomon effect through bootstrap resampling, which allows us to explore the set of near-optimal models by accounting for the inherent variance in the data \citep{biecek2024position}. This approach moves the focus from an arbitrary performance threshold to the stability of model behavior across different realizations of the training data.

The bootstrap-based Rashomon set can be formally defined as follows. Let $D = \{(x_i, y_i)\}_{i=1}^n$ be the original dataset. We generate $B$ bootstrap samples $D^*_1, D^*_2, \dots, D^*_B$ by sampling from $D$ with replacement. For each sample $D^*_b$, an empirical risk minimizer (a model) $f_b$ is trained:\begin{equation}f_b = \arg\min_{f \in \mathcal{F}} \hat{L}(f, D^*_b)\end{equation}The resulting collection of models constitutes the bootstrap Rashomon set:\begin{equation}\hat{R}_{B} = { f_1, f_2, \dots, f_B }.\end{equation}In this framework, each $f_b \in \hat{R}_{B}$ represents a candidate explanation of the underlying phenomenon that is optimal under a specific perturbation of the data. While these models achieve comparable predictive accuracy on the population level, they may differ in their feature dependencies and decision rules.

This interpretation ambiguity necessitates robust methodologies that move beyond single-model explanations. By analyzing the collective behavior of $\hat{R}_{B}$—a process we refer to as \textit{rashomonizing} the analysis—we can quantify the stability of our findings. If a predictive relationship (e.g., the effect of age on anxiety) remains consistent across the entire set $\hat{R}_{B}$, it provides strong evidence of a robust biological or psychological trend. Conversely, significant divergence among models in the set indicates that the data is underspecified for that particular region, highlighting the epistemic uncertainty inherent in the modeling process \citep{cavus2025beyond, biecek2024performance}.

There are several ways, such as using different seeds in data splitting \citep{ganesh2024cost}, various hyperparameter settings \citep{cavus2025role}, different model families, pre-trained models \citep{renard2024understanding}, AutoML tools \citep{cavus2025investigating}, or any combination of these, to create the Rashomon set. However, models within this set may differ substantially in terms of variable importance \citep{dong2020exploring, donnelly2023rashomon}, decision boundaries, and inference outcomes \citep{marx2020predictive}. This interpretation ambiguity necessitates robust methodologies that move beyond single-model explanations to quantify the collective behavior of the entire Rashomon set, which is referred to as the Rashomon perspective \citep{cavus2025beyond, biecek2024performance, biecek2024position}.
In this paper, we operationalize this perspective by rashomonizing the interpretation of predictive relationships through partial dependence profiles computed over multiple well-performing models, rather than relying on explanations derived from a single fitted model.

Although the Rashomon perspective may appear conceptually related to ensemble learning, the two approaches differ in both objective and scope. Ensemble learning primarily aims to improve predictive performance by aggregating predictions from a potentially large and heterogeneous collection of models, often without regard to their individual explanatory quality. In contrast, the Rashomon perspective explicitly restricts attention to a subset of near-optimal, high-performing models—the Rashomon set—and focuses on comparing and synthesizing their explanatory behaviors rather than their predictions. This restriction is motivated by the expectation that models achieving similarly strong performance are more likely to provide reliable and meaningful explanations, whereas including weaker or poorly fitting models, as in standard ensembles, may dilute or obscure interpretability. Consequently, the Rashomon approach can be viewed as operating on a principled subset of the ensemble space, prioritizing interpretability and epistemic robustness over predictive aggregation.

\section{Partial Dependence Profiles}\label{sec3}

Partial Dependence Profiles (PDP) \citep{friedman2001greedy} is an explainable artificial intelligence tool that provides the marginal effect of a specific predictor on the predicted outcome, allowing for interpretations of the complex, non-linear patterns learned by black-box models.

Formally, let the input vector $x$ be partitioned into two sets of variables: $x_j$, the feature of interest, and $x_{-j}$, which represents all other features in the dataset. For a model $f$, the partial dependence of the prediction on the feature $x_j$ is defined as:

\begin{equation}
    f_j(x_j) = \mathbb{E}_{X_{-j}} [f(x_j, X_{-j})].
\end{equation}

\noindent Since the true distribution of $X_{-j}$ is typically unknown, this expectation is estimated by calculating the average of the model's predictions over the observed data $D = \{(x_i, y_i)\}_{i=1}^n$:

\begin{equation}
    \hat{f}_j(x_j) = \frac{1}{n} \sum_{i=1}^n f(x_j, x_{i,-j}),
\end{equation}

\noindent where $x_{i,-j}$ are the actual values of the remaining features for the $i$-th observation in the dataset. 

By applying this estimation to every model within the Rashomon set, we \textit{rashomonize} the interpretation of predictor effects, allowing us to assess the robustness of the discovered relationships. If the resulting profiles are consistent across the entire set, it provides strong evidence that the effect of feature $x_j$ is a stable characteristic of the underlying phenomenon rather than an artifact of a specific model's architecture. Conversely, significant divergence among the profiles indicates regions of the feature space where the data is underspecified, leading to the derivation of robust insights about the relations from the model.

\section{PHQ4}\label{sec4}

The Patient Health Questionnaire-4 (PHQ-4) is an ultra-brief, four-item self-report questionnaire designed for screening depressive and anxiety symptoms \citep{kroenke_ultra-brief_2009}. It is widely used in both clinical and general population settings due to its brevity and effectiveness \citep{lowe_4-item_2010, kazlauskas_psychometric_2023, caro-fuentes_systematic_2024}. It was developed by combining two existing ultra-brief screeners: the first two-item Patient Health Questionnaire (PHQ-2) for depression \citep{kroenke_patient_2003} and the first two-item Generalized Anxiety Disorder Scale (GAD-2) for anxiety \citep{spitzer_brief_2006}. The PHQ-2 itself is derived from the nine-item PHQ-9, focusing on the core symptoms of depressed mood and loss of interest \citep{kroenke_patient_2003}. Similarly, the GAD-2 is a short version of the seven-item GAD-7, addressing feeling nervous/anxious and inability to control worrying, which are core symptoms of generalized anxiety disorder \citep{spitzer_brief_2006}.

The PHQ-4 instrument begins with the stem question: ``Over the last 2 weeks, how often have you been bothered by the following problems?''. Responses are scored on a 4-point Likert scale, ranging from 0 (``not at all'') to 3 (``nearly every day'') \citep{kroenke_ultra-brief_2009}. This scoring results in a total PHQ-4 score ranging from 0 to 12, while the PHQ-2 and GAD-2 subscale scores each range from 0 to 6 \citep{kroenke_patient_2003, spitzer_brief_2006, kroenke_ultra-brief_2009}. The developers suggested a cut-off score of $\geq 3$ for both the PHQ-2 and GAD-2 to indicate probable cases of depression or anxiety, respectively. A score of $\geq 6$ on the total PHQ-4 can indicate distress. It's important to note that the PHQ-4 is a screening tool and not a clinical diagnostic instrument \citep{kroenke_ultra-brief_2009}.

The initial validation of the PHQ-4 was conducted using data from 2,149 patients across 15 primary care clinics in the United States \citep{kroenke_ultra-brief_2009}. Subsequent studies have further validated and standardized the PHQ-4 in large general population samples \citep{lowe_4-item_2010}. Its development aims to provide a reliable and valid tool for efficient mass screening, particularly valuable in busy clinical settings or for large-scale epidemiological studies where time and resources are limited \citep{adzrago_reliability_2024}.

The PHQ-4 has demonstrated strong psychometric properties across diverse populations and settings. Multiple studies consistently support a two-factor structure for the PHQ-4, with distinct depression (PHQ-2) and anxiety (GAD-2) factors \citep{mendoza_factor_2022}. The PHQ-4 and its subscales (PHQ-2 and GAD-2) have also demonstrated high internal consistency. Cronbach's alpha values typically range from 0.78 to 0.92 for the overall PHQ-4, 0.65 to 0.86 for PHQ-2, and 0.74 to 0.90 for GAD-2 \citep{adzrago_reliability_2024}. McDonald's omega also supports these findings, with reported values of 0.85 for PHQ-4, 0.77 for PHQ-2, and 0.78 for GAD-2 in recent studies \citep{kazlauskas_psychometric_2023}. Its construct validity is supported by the expected intercorrelations with other related self-report measures and demographic risk factors \citep{lowe_4-item_2010}, and its convergent validity is demonstrated by significant positive correlations with other scales measuring psychological distress, anxiety, and depression. Examples include positive correlations with stress (DASS-S), negative affect (PANAS-N), Brief Symptom Inventory (BSI), Hospital Anxiety and Depression Scale (HADS), PHQ-9, Hamilton Anxiety Rating Scale (HARS), Beck Anxiety Inventory (BAI), Hamilton Depression Rating Scale (HDRS), and Penn State Worry Questionnaire (PSWQ) \citep{caro-fuentes_systematic_2024}. 

PHQ-4 scores and the prevalence of anxiety and depression symptoms are consistently associated with various sociodemographic characteristics such as  \textit{Age}: Younger adults (e.g., 18-34 years) tend to report higher scores for depression and anxiety than older adults \citep{caro-fuentes_systematic_2024}; \textit{Marital Status}: Individuals who are divorced or single tend to score higher on the PHQ-4 and its subscales compared to those who are married \citep{adzrago_reliability_2024}; \textit{Employment Status}: Unemployed individuals consistently report significantly higher scores for depression and anxiety symptoms than employed individuals \citep{adzrago_reliability_2024}; \textit{Education Level}: There is an observed trend that individuals with lower educational levels tend to have higher PHQ-4 score \citep{caro-fuentes_systematic_2024}; \textit{Household Income}: Lower household income is also associated with higher levels of depression and anxiety symptoms as measured by the PHQ-4 \citep{adzrago_reliability_2024}.

Overall, the PHQ-4 is considered a robust, reliable, and valid ultra-brief instrument for screening depression and anxiety symptoms across a wide range of populations and contexts when compared with the full versions of the PHQ-9 and GAD-7. Its consistent two-factor structure, high internal consistency, and established construct validity make it a valuable tool for identifying individuals who may require further assessment, thereby contributing to the optimization of healthcare resources \citep{wicke_update_2022}.

\section{Experiments}\label{sec5}

\subsection{Dataset}\label{sec5.1}

The dataset consists of data collected between March 2020 and June 2024 through online recruitment via the official Instagram and Facebook pages of a university-based research laboratory that specializes in public mental health. Table~\ref{tab:dataset} presents detailed descriptions of the variables used in modeling.
The final sample consisted of 34,646 respondents, all of whom were 18 years of age or older. All data collection procedures adhered to both Brazilian and international standards for research involving human participants. Electronic informed consent was secured from all participants before their involvement in the survey. Approval was granted by Brazil’s National Council for Ethics in Research (Conselho Nacional de Ética em Pesquisa [CONEP], under protocol number 30485420.6.0000.0008).

\begin{table*}[h]%
    \centering
    \caption{Descriptive Statistics of Categorical Variables, Brazil (2020-2024)}
    \label{tab:dataset}
    {\small
    \begin{tabular}{p{3cm} p{4cm} p{1.5cm} p{1.5cm}}\toprule
Variable & Level &	N(34443)	&	F\%	\\
\hline	
Year	&		&		&		\\
	&	2020	&	8024	&	23.3	\\
	&	2021	&	1540	&	4.5	\\
	&	2022	&	2208	&	6.4	\\
	&	2023	&	2246	&	6.5	\\
	&	2024	&	20425	&	59.3	\\
\hline	
Sex	&		&		&		\\
	&	Male	&	5662	&	16.4	\\
	&	Female	&	28781	&	83.6	\\
\hline	
Age strata	&		&		&		\\
	&	18-30	&	4685	&	13.6	\\
	&	31-40	&	6557	&	19.0	\\
	&	41-50	&	10548	&	30.6	\\
	&	51-60	&	10583	&	30.7	\\
	&	+61	&	2070	&	6.0	\\
\hline	
Education level	&		&		&		\\
	&	Bellow High school	&	8154	&	23.7	\\
	&	High school complete	&	8736	&	25.4	\\
	&	Graduate	&	17553	&	51.0	\\
\hline	
Hour of day	&		&		&		\\
	&	00-05: Night/Early	&	5072	&	14.7	\\
	&	06-11: Morning 	&	4918	&	14.3	\\
	&	12-16: Afternoon/Lunch    	&	8775	&	25.5	\\
	&	17-20: Evening/Peak  	&	8381	&	24.3	\\
	&	    21-23: Late Night	&	7297	&	21.2	\\
\hline	
Weekday \ weekend	&		&		&		\\
	&	Weekday	&	26186	&	76.0	\\
	&	Weekend	&	8257	&	24.0	\\
\hline	
GAD-2 Scores	&		&		&		\\
	&	0	&	1816	&	5.3	\\
	&	1	&	2388	&	6.9	\\
	&	2	&	7483	&	21.7	\\
	&	3	&	3540	&	10.3	\\
	&	4	&	5423	&	15.7	\\
	&	5	&	4077	&	11.8	\\
	&	6	&	9716	&	28.2	\\
\hline
PHQ-2 Scores	&		&		&		\\
	&	0	&	3629	&	10.5	\\
	&	1	&	3199	&	9.3	\\
	&	2	&	7923	&	23.0	\\
	&	3	&	3402	&	9.9	\\
	&	4	&	4617	&	13.4	\\
	&	5	&	3332	&	9.7	\\
	&	6	&	8341	&	24.2	\\
\hline
    
    \end{tabular}
    }
\end{table*}

\subsection{Design}\label{sec5.2}

This study employed a methodology for model interpretation based on PDPs. They were utilized to comparatively examine how a Random Forest model's output responds to a specific variable across different subgroups. This approach makes it possible to visualize the model's reaction to variables through group-level average predictions. The analysis focused on understanding the model's behavior by comparing PDPs of the trained model on response variables PHQ-2 and GAD-2, for independent variables including \texttt{age}, \texttt{sex}, \texttt{education}, \texttt{hour}, and \texttt{weekday\_weekend}. We followed the Rashomon approach given in Algorithm~\ref{alg:rashomon}. 

\begin{algorithm}[h!]
\caption{PDPs with Rashomon Perspective}
\label{alg:rashomon}
\begin{algorithmic}[1]
\Require Dataset $D$, Iterations $B = 100$, Target feature $x_j$, Grouping variable $g$.
\Ensure Aggregated PDPs with $95\%$ confidence intervals.
\For{$b \gets 1$ \textbf{to} $B$}
    \State $D_b \gets$ Generate a bootstrap sample from $D$ with replacement.
    \State $f_b \gets$ Train a Random Forest model on $D_b$.
    \State $\hat{f}_{j,g,b} \gets$ Calculate grouped PDP for $x_j$ stratified by $g$.
\EndFor
\State $\mathcal{R} \gets \{ \hat{f}_{j,g,1}, \dots, \hat{f}_{j,g,B} \}$ \Comment{Collection of profiles from the bootstrap Rashomon set $\hat{R}_{B}$}
\For{each unique value in range of $x_j$ and group $g$}
    \State Compute average prediction: $\mu = \text{mean}(\hat{f}_{j,g,b})$.
    \State Compute $2.5\%$ and $97.5\%$ quantiles for the confidence intervals.
\EndFor
\State \Return PDPs and their confidence intervals.
\end{algorithmic}
\end{algorithm}

This involved training multiple models with similar predictive performance and comparing their behaviors. Specifically, bootstrap samples ($N = 100$) were drawn from the original dataset. For each sample, a separate Random Forest model was trained to predict the response. For each of these models, a corresponding PDP for the \texttt{age} variable was generated for groups based on \texttt{sex}, \texttt{education}, and \texttt{weekday\_weekend}. These multiple profiles were used to construct PDPs with bootstrap confidence intervals, which were then visualized as confidence bands around the PDP. The resulting bootstrap confidence intervals quantify the variability of the estimated PDP due to sampling uncertainty and model instability, capturing how sensitive the estimated marginal effect of age is to perturbations in the training data rather than uncertainty in individual predictions. Narrow bands indicate stable effects across resampled datasets, while wider bands suggest greater uncertainty in the estimated relationship. This approach provides a clear representation of the variability and reliability of the model's predictions, directly addressing the interpretation ambiguity inherent in the Rashomon Set.

\subsection{Results}\label{sec5.1}

\begin{figure*}[t]
  \centering
    \includegraphics[width=\linewidth]{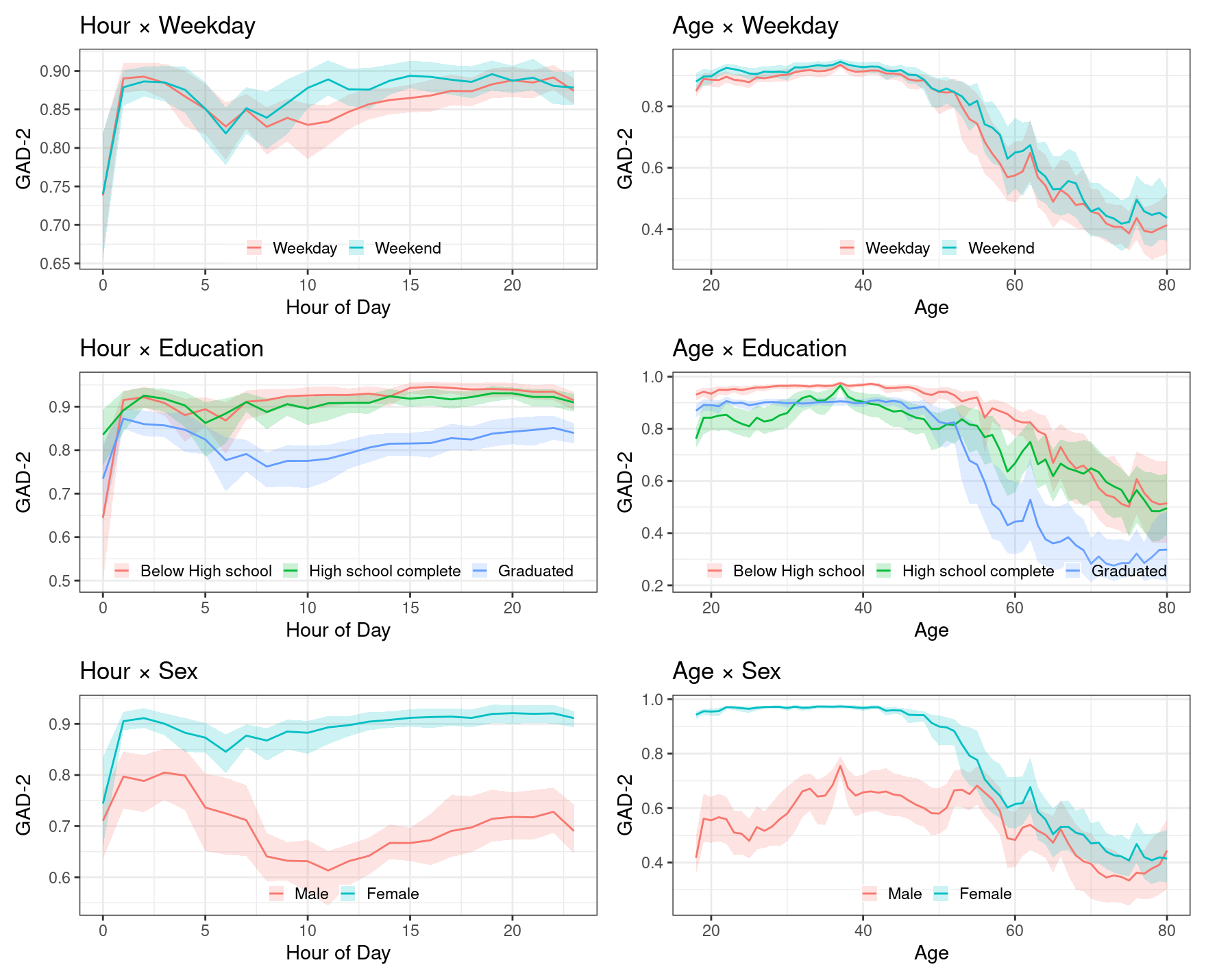}
    \caption{The Grouped PDPs for GAD-2 Risk Across the Factors. They illustrate the model-predicted relationship between the plotted factors and the estimated probability of the GAD-2 score being greater than 3. The lines represent the predicted probability values, and the surrounding bands indicate the 95\% confidence intervals. Non-overlapping confidence intervals between groups suggest differences in predicted risk.}
    \label{fig:GAD-2}
\end{figure*}

\begin{figure*}
 \centering
    \includegraphics[width=\linewidth]{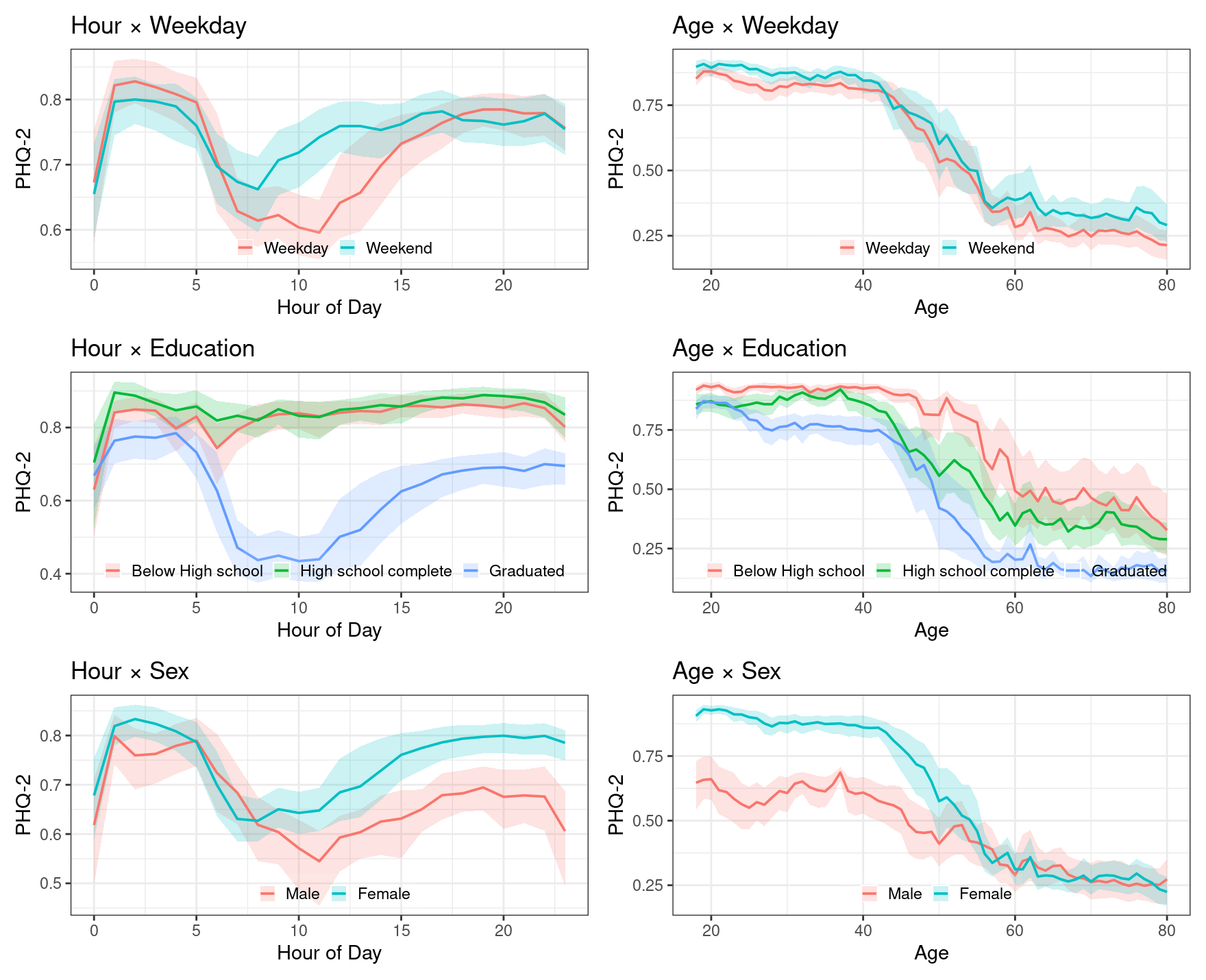}
    \caption{The Grouped PDPs for the PHQ-2 Risk Across the Factors. They illustrate the model-predicted relationship between the plotted factors and the estimated probability of the PHQ-2 score being greater than 3. The lines represent the predicted probability values, and the surrounding bands indicate the 95\% confidence intervals. Non-overlapping confidence intervals between groups suggest differences in predicted risk.}
    \label{fig:PHQ-2}
\end{figure*}

The following grouped Rashomon PDPs illustrate the model-predicted relationship between the variables \texttt{hour}, \texttt{age}, \texttt{weekday}, \texttt{education}, and \texttt{sex}, and the estimated probability of GAD-2 and PHQ-2 scores being greater than 3. The Figures~\ref{fig:GAD-2} and~\ref{fig:PHQ-2} reveal how the predicted risk probability changes with variations in the considered variable on the x-axis. The lines represent the predicted probability values, while the surrounding bands indicate the $95\%$ confidence intervals. Crucially, when the confidence intervals for two distinct groups do not overlap, it provides strong evidence that a difference exists. These profiles serve as a foundation for understanding the extent to which demographic and temporal factors influence the risk of anxiety and depression.

The grouped Rashomon PDPs for GAD-2, as shown in Figure~\ref{fig:GAD-2}, reveal a pronounced temporal pattern in anxiety-symptom risk across the day. For both weekdays and weekends, predicted probabilities peak sharply during the early morning hours, particularly around 1 to 3 AM, before stabilizing at moderately elevated levels throughout the remainder of the day. Weekend predictions remain consistently above weekday predictions, indicating slightly heightened anxiety risk on weekends across nearly all hours. In the age–weekday interaction, predicted GAD-2 risk is relatively stable and high from early adulthood through approximately midlife, followed by a steady decline beginning around age 50 and continuing into older adulthood. Similar to the hour-based patterns, weekend predictions exceed weekday predictions across much of the age range, with the difference becoming more pronounced in late middle age.

Educational differences also manifest clearly in the GAD-2 in terms of PDPs. When anxiety risk is examined across hours of the day, all education groups show the characteristic early-morning spike, but the lowest-education group consistently exhibits the highest predicted anxiety probabilities throughout the entire day. The highest-education group, by contrast, displays substantially lower predicted probabilities, especially in the late-night and early-morning hours where group differences are most pronounced. A similar gradient appears across age: while the general pattern of stable early-adulthood risk followed by age-related decline emerges in all groups, the lowest-education category maintains the highest predicted risk across nearly all ages. The highest-education group shows considerably lower predicted anxiety probabilities, particularly after age 50, accompanied by widening confidence intervals at older ages, likely reflecting reduced sample density.

Sex-based profiles also display systematic and robust differences in predicted GAD-2 risk. Across the day, both sex groups demonstrate the early-morning peak in predicted anxiety symptoms. Still, one group maintains consistently higher risk at all hours, with the divergence persisting through the afternoon and evening. When examined across age, the same group exhibits elevated predicted probabilities throughout early and mid-adulthood, followed by a decline that mirrors the overall population trend. Although both groups show decreasing anxiety-symptom risk after midlife, the gap between them remains apparent through older ages. Taken together, these PDPs suggest that while hour of day and weekday/weekend status shape the temporal contour of predicted anxiety risk, demographic characteristics—especially education and sex—exert strong, consistent shifts in overall risk levels. Age-related declines in predicted GAD-2 risk appear robust across all groups, highlighting a common life-course trajectory in anxiety-symptom probability.

The grouped PDPs for PHQ-2, as given in Figure~\ref{fig:PHQ-2}, show several noteworthy temporal patterns in predicted PHQ-2 risk. Across both weekdays and weekends, the probability of scoring above the PHQ-2 threshold peaks in the early morning hours, particularly around 1 to 3 AM, declines sharply toward the morning, and gradually rises again throughout the afternoon and evening. Weekend predictions are consistently higher than weekday predictions across most hours of the day, suggesting that individuals may experience slightly elevated depressive-symptom risk during weekends. When age is examined in conjunction with weekday status, the model indicates that younger adults exhibit the highest predicted risk, with a steady decline as age increases and a flattening at low levels after roughly age 60. Similar to the hour-based profiles, weekend predictions remain slightly higher than weekday predictions across a broad age range, particularly from young adulthood through midlife.

Educational differences emerge clearly in the grouped PDPs. When considered alongside hour of day, the highest-education group consistently shows substantially lower predicted PHQ-2 probabilities relative to the lower-education groups, despite all groups following a similar diurnal shape marked by early-morning peaks and late-morning troughs. A similar pattern is observed across age: although all education groups exhibit a decline in predicted risk from early adulthood to old age, individuals in the highest-education category maintain markedly lower predicted probabilities across the entire age spectrum. The lowest-education group exhibits the highest predicted risk across most ages, particularly throughout midlife, with wider uncertainty at older ages where data are likely sparser.

Sex differences also appear stable and robust across both temporal dimensions. When examined by hour of day, both sex groups follow a similar daily rhythm, but one sex group shows consistently higher predicted risk at nearly all hours. This pattern persists across age as well: predicted PHQ-2 risk declines steadily with increasing age for both groups, yet one group maintains a higher probability across most of the lifespan. The difference narrows somewhat at older ages, though the overall trend remains consistent. Taken together, these PDPs indicate that while temporal factors such as hour of day and weekday versus weekend exert cyclical and situational influences on model predictions, demographic characteristics—particularly age, education, and sex—produce stable and meaningful shifts in predicted depressive-symptom risk.

\section{Discussion}\label{sec6}

This paper investigates the application of the Rashomon effect to two ultra-brief self-report questionnaires, utilizing a substantial sample size exceeding 34,000 responses. The Rashomon effect posits that multiple distinct predictive models can achieve comparable accuracy, thus mandating that researchers analyze the entire ``Rashomon set'' of plausible models rather than concentrating on a singular one.

To substantiate this concept, the authors employed machine learning on a large dataset of scores derived from the Patient Health Questionnaire-4 (PHQ-4), a widely adopted, ultra-brief screening instrument for symptoms of anxiety (GAD-2) and depression (PHQ-2). The methodology entailed the combination of a Random Forest model with bootstrap analysis to generate partial dependence profiles, which were subsequently used to evaluate the robustness and reliability of predictions across diverse subgroups.

The robust experimental findings demonstrate the influence of various factors on the predicted risk probability. Overall, the results indicate that demographic variables (e.g., age, educational attainment, and sex) induce consistent, structural shifts in anxiety and depression risk levels. Conversely, temporal variables, like the hour of the day, led to predictable cyclical fluctuations. A negative correlation was observed between age and the risk of anxiety or depression; increasing age corresponded with lower risk. Additionally, GAD-2 and, especially, PHQ-2 scores exhibited fluctuations, notably increasing during the night or towards the end of the day.

These findings align with the existing literature, which frequently reports gender \citep{su_gender_2024, shawon_gender_2024}, age \citep{hobbs_relationship_2014}, and education level \citep{sousa_reliability_2015} differences in scores on the GAD and PHQ. However, the influence of the hour of the day and whether it is a weekday or weekend represents a novel finding, previously unreported in the literature. Differences in anxiety levels associated with physical activity patterns and excessive daytime arousal have been noted in some studies \citep{pastre_actigraphy_2022}. However, the potential influence of this specific pattern on self-report scales like the GAD has not been examined in the literature. Therefore, this study offers new perspectives by suggesting that the time of day and even the day of the week may affect responses on such scales.

From a public mental health perspective, our findings can impact population monitoring, particularly for large-scale screening data, and they can be interpreted. To identify stable demographic gradients, alongside systematic circadian and circaseptan fluctuations, helps the understanding that symptom prevalence estimates based on brief instruments, such as the PHQ-4, may be partly shaped by the timing of data collection; an issue that has become increasingly visible in recent mental health surveillance efforts \citep{Walther2025Depressive}. In large and regular population surveys, taking into account temporal patterns may introduce small but systematic sources of bias, particularly when prevalence estimates are compared across survey waves, regions, or population subgroups \citep{vigo_true_2022, patel_lancet_2018}.

\section{Limitations}\label{sec7}

A primary limitation of the current study is the unbalanced sample, which consists predominantly of female participants. Furthermore, the use of digital convenience sampling for recruitment, while generating a large number of participants, prevents an accurate representation of the Brazilian general population. The latter, plus the unbalanced sample, limits the generalizability of the findings, especially when considering male populations. However, it is important to note that this limitation is less significant for variables concerning circadian and circaseptan rhythms, which are adequately represented within the current dataset (see Table \ref{tab:dataset}).

Beyond the limitations of sample generalizability, this study acknowledges inherent methodological constraints associated with the Rashomon perspective. First, PDPs capture the conditional relations; thus, no correlations between the variables are considered. This may be problematic when the correlations between variables in complex datasets. Second, while the Rashomon approach provides stability, it is important to note the computational cost involved in training separate models to create the Rashomon set, which can be prohibitive in larger-scale or high-dimensional applications. Third, the operationalization of the Rashomon set through bootstrap resampling depends on the number of iterations and the diversity of the generated models. While this approach avoids the use of an arbitrary performance threshold, the scope of the multiplicity analysis is inherently linked to the computational scale of the resampling process. Although $100$ iterations is sufficient for stabilizing the partial dependence profiles, it is possible that the resulting Rashomon set captures a representative distribution rather than the exhaustive space of all near-optimal models, potentially leaving some rare but plausible structural narratives unexplored. Furthermore, while our current approach focuses on data-driven perturbations via bootstrapping, future work could expand the Rashomon set by perturbing the optimization process itself—such as using different random seeds or hyperparameter settings—or by utilizing AutoML frameworks to explore a broader range of model architectures. Such extensions would allow for a more exhaustive exploration of the Rashomon set, addressing both data-related and algorithmic sources of multiplicity.

\section{Conclusion}\label{sec8}

Continuous screening for common mental health disorders, such as anxiety and depression, coupled with advanced techniques like the Rashomon effect for selecting the optimal machine learning model, allows for the extraction of significant information from large datasets. This approach emphasizes the value of continuous monitoring while also highlighting the need to explore novel methods for data analysis and dissemination. Utilizing a substantial sample from two ultra-brief self-report instruments, we specifically underscore that variability associated with circadian and circaseptan rhythms necessitates meticulous consideration of the time of day and the specific day on which these scales were administered. By rashomonizing the analysis of large-scale PHQ-4 screening data, this paper demonstrates how demographic gradients and temporal rhythms can be interpreted more robustly when model multiplicity is explicitly acknowledged.

\section*{Acknowledgments}
AF is a recipient of a Research Fellowship from the National Council for Scientific and Technological Development (Conselho Nacional de Desenvolvimento Científico e Tecnológico – CNPq, Brazil) and the Coordination for the Improvement of Higher Education Personnel (Coordenação de Aperfeiçoamento de Pessoal de Nível Superior – CAPES, Brazil). The author acknowledges the support provided by these institutions, which contributed to the development of this research. The funders had no involvement in the study design, data collection, analysis, interpretation, or the writing of the manuscript.

\section*{Reproducibility Statement}
The implementation code, preprocessing scripts, and reproducibility documentation for this paper are available at \url{https://github.com/mcavs/Analyzing-Temporal-Factors-for-Anxiety-and-Depression-Symptoms-with-the-Rashomon-Perspective}.

\bibliography{sn-bibliography}

\end{document}